\documentclass[12pt]{article}
\usepackage{amsfonts,amssymb,amsmath,graphicx,epic}
\newcommand{\nn}{\nonumber}
\newcommand{\ZZ}{\mathbb{Z}}
\newcommand{\RR}{\mathbb{R}}
\newcommand{\CC}{\mathbb{C}}
\newcommand{\N}{{\cal N}}
\def\be{\begin{equation}}
\def\ee{\end{equation}}
\def\bea{\begin{eqnarray}}
\def\eea{\end{eqnarray}}
\def\bz{\bar{z}}

\newcommand{\fft}[2]{{\frac{#1}{#2}}}

\begin{document}

\thispagestyle{empty}
\renewcommand{\thefootnote}{\fnsymbol{footnote}}

{\hfill \parbox{3.5cm}{
        HUTP-05/A0023 \\ MCTP-05--79   \\
}}

\bigskip\bigskip

\begin{center} \noindent \Large \bf
The D3/D7 Background and \\
Flavor Dependence of Regge Trajectories    
\end{center}

\bigskip\bigskip\bigskip

\begin{large}
\centerline{  
Ingo Kirsch${}^a$ and Diana Vaman${}^b$ }
\end{large}

\bigskip\bigskip
\begin{center}

${}^a$ \textit{Jefferson Laboratory of Physics, Harvard University, \\
  Cambridge, MA 02138, USA}\\
\medskip

${}^b$  \textit{Michigan Center for Theoretical Physics, Randall Laboratory
of Physics,\\
The University of Michigan, Ann Arbor, MI 48109--1120}
\end{center}

\bigskip
\bigskip\bigskip

\bigskip\bigskip

\renewcommand{\thefootnote}{\arabic{footnote}}

\centerline{\bf \small Abstract}
\medskip

{\small In the context of AdS/CFT with flavor, we consider the type
IIB supergravity solution corresponding to a fully localized D3/D7
intersection.  We complete the standard metric ansatz by providing an
analytic expression for the warp factor, under the assumption of a
logarithmically running axion-dilaton. From the gauge dual perspective,
this behavior is related to the positive beta function of the ${\cal
N}=4$, $d=4$ $SU(N_c)$ super Yang-Mills gauge theory, coupled to
$N_f$~fundamental ${\cal N}=2$ hypermultiplets.  We comment on the
existence of tadpoles and relate them to the same gauge theory beta
function. Next we consider a classical spinning string configuration
in the decoupling limit of the D3/D7 geometry and extract the flavor
($N_f$) dependence of the associated meson Regge trajectory.
Including the backreaction of the D7-branes in the supergravity dual
allows for going beyond the quenched approximation on the dual gauge
theory side.  }

\newpage
\section{Introduction}

In the recent past, much effort has gone into studying the AdS/CFT
duality of Yang-Mills theories with matter in the fundamental
representation of the gauge group. Starting with the work of Karch and
Katz \cite{Karch}, fundamental matter was introduced as a new open
string sector arising through the embedding of probe branes into
various supergravity backgrounds
\cite{Kruczenski:2003be}--\cite{Apreda:2005yz}. This approach
corresponds to the quenched approximation in lattice QCD which ignores
the effect of the creation and annihilation of virtual quark-antiquark pairs.

In a parallel development, supergravity backgrounds of fully localized
Dp/D(p+4) brane intersections have been constructed which include the
backreaction of the ``flavor'' D(p+4) brane \cite{Cherkis,
Nastase:2003dd, Kehagias:1998gn, afm, Grana, Burrington,
Liu:2004ru}. Such brane configurations allow for the discussion of
flavors beyond the quenched approximation. An interesting example is
the D3/D7 brane intersection which, in contrast to the same system in
the probe approximation \cite{Karch}, includes the backreaction of the
D7-branes on the geometry. This configuration describes an $\N=2$
gauge theory obtained from the coupling of ${\cal N}=4$ $SU(N_c)$
super Yang-Mills theory to $N_f$ hypermultiplets in the fundamental
representation of the gauge group.
Similar to QED and $\phi^4$~theory, the theory has a positive beta
function leading to an ultraviolet divergence in the gauge coupling
constant.  An ultraviolet Landau pole usually indicates the breakdown
of the perturbative field theory at high energies, and it is not yet
understood how this pathology can be cured.

In this paper we complete the construction of the type IIB
supergravity solution for the D3/D7 intersection restricted to the
near-core region of the D7-branes, where the axion-dilaton has a simple
logarithmic behavior. Unlike the full complex structure of the
D7-branes, which is finite everywhere, the dilaton runs into a
singularity at some finite distance away from the D7-branes.

As we will show, there is a one-to-one map between the logarithmic dilaton
profile and the one-loop running of the gauge coupling. This precisely
maps the Landau pole in the field theory to the dilaton divergence in
the supergravity solution. Similarly, the chiral anomaly of the
$U(1)_{\cal R}$ symmetry, which leads to a non-vanishing theta angle,
is appropriately represented by a non-trivial axion field.

As shown in \cite{afm, Grana}, the construction of a supergravity
solution for the D3/D7 system, reduces to the problem of finding a
solution to a Poisson equation for the warp factor. Despite claims to
the contrary, it is possible to find an analytical solution for the
warp factor.  We show that the Poisson equation can be reduced to a
linear ordinary differential equation which can be solved by means of
techniques developed by Gesztesy and \mbox{Pittner} \cite{Gesztesy}. The
general solution is represented by a uniformly convergent series
expansion for which we give an explicit proof of convergence.

Due to the logarithmic ultraviolet divergence, the string background
is not finite and is expected to have an uncanceled tadpole.
Nevertheless, as pointed out in~\cite{Leigh:1998hj},
if the tadpole is logarithmic, the background is
still consistent as far as the embedding of the non-conformal gauge
theory is concerned. Logarithmic tadpoles do not reflect gauge
anomalies, but instead provide the correct one-loop running of the
gauge coupling. Using results found in \cite{DiVecchia:2005vm}, we
will show that the open string one-loop annulus amplitude has indeed
the correct logarithmic behavior expected from the positive beta
function of the $\N=2$ field theory.

The universal structure of the D3/D7 configuration gives rise to many
applications in string theory. Apart from its application to strong
interactions, the D3/D7 system has also received renewed attention in
a cosmological context. A prominent example is the D3/D7 model in
\cite{Dasgupta:2002ew} which has an effective description as hybrid
inflation. It might even turn out that D3 and D7-branes are the
preferred branes in brane cosmology, as can be argued by a discussion
of the decay and annihilation rates of D-branes in a higher-dimensional
universe~\cite{Lisa}.  Finally, the D3/D7 plays an essential role in
the construction of semi-realistic four-dimensional string vacua
\cite{Lust:2005bd}.

In the second part of the paper, we study meson Regge trajectories
in the $\N=2$ gauge theory dual to the previously constructed D3/D7
supergravity solution. We are particularly interested in the
dependence of the meson spectrum on the number of sea
quark flavors introduced by the D7-branes.

Following the analysis in \cite{Kruczenski:2003be, Kruczenski:2004me,
Paredes:2004is}, we determine the spectrum of mesons with large
spin~$J$  by computing the energy and angular
momentum of a semi-classical string rotating in the near-horizon
geometry of the D3/D7 intersection.  This string is attached to an
additional probe D7-brane which corresponds to a massive flavor~$Q$ in
the background of $N_f$ massless flavors $q_i$. The setup is similar
to the one in~\cite{Kruczenski:2003be}, but with $N_f$ flavors of
light sea quarks turned on.

By varying the number of D7-branes $N_f$ and keeping the mass of the
single flavor $Q$ fixed, we determine the Regge trajectory
of the meson $Q \bar Q$. We find that for a meson with a particular spin
$J$, the energy as well as the string tension decreases with the
number of massless flavors, whereas the string length, which measures
the distance between the quarks inside the meson, remains unaffected
by the sea quark flavors.

In the field theory, the decrease in the string tension can be
ascribed to the influence of virtual color-singlet $q\bar q$ pairs on
the color flux inside the meson $Q\bar Q$. As it is well-known,
virtual $q\bar q$ pairs cause a screening of the color charge,
effectively diminishing the color force and with it the string
tension. This quark-loop effect amounts to going to first order
in $N_f/N_c$ beyond the quenched approximation in the gauge theory and
arises as a consequence of including the backreaction of the D7-branes
in the D3/D7 supergravity background. According to the standard lore
of AdS/CFT, a gauge theory one-loop quantum effect is captured at the
classical level by the supergravity dual.

The paper is organized as follows. In Sec.~\ref{sec2}, we revisit the
D3/D7 configuration and derive an analytical expression for the warp
factor in terms of a uniformly convergent series expansion. From this
we also deduce the supergravity solution for fractional D3/D7 branes
which describes $\N=2$ $SU(N_c)$ super Yang-Mills theory with
$N_f$ fundamental hypermultiplets, but without an adjoint
hypermultiplet. We further discuss several issues related to the
consistency of the string theory embedding of a non-conformal gauge
theory with running gauge coupling.  In~Sec.~\ref{sec3} we discuss the
flavor dependence of Regge trajectories in the D3/D7 brane
theory. A~summary of the D3/D7 supergravity solution can be found in
the conclusions in Sec.~\ref{sec4}.  App.~\ref{appD7} contains a
review on the D7-branes geometry.  App.~\ref{appGP} contains the
convergence proof of the series expansion of the warp factor.

\section{The fully backreacted D3/D7 solution} \label{sec2}
\setcounter{equation}{0}\setcounter{figure}{0}\setcounter{table}{0}

In the following we construct the supergravity solution for a fully
localized D3/D7 intersection in flat space. The corresponding
near-horizon limit of this solution is dual to ${\cal N}=2$ super
Yang-Mills theory with $N_f$ hypermultiplets in the fundamental and
one hypermultiplet in the adjoint representation of the gauge group
$SU(N_c)$. This set-up is T-dual to the D2/D6 and D4/D8 geometry
derived in \cite{Cherkis} and \cite{Nastase:2003dd},
respectively. Previous work on the D3/D7 solution in flat space can be
found in \cite{Kehagias:1998gn, afm, Grana, Burrington}.

\subsection{The D3/D7 solution}

The D3/D7 brane intersection consists of a stack of $N_c$ coincident
D3-branes which is embedded into the world volume of $N_f$
D7-branes. The D7-branes are located in ten-dimensional flat space
such that they share four longitudinal directions with the D3-branes
while wrapping or spanning four out of the six transverse
directions. This embedding may be represented pictorially as
\begin{center}
\begin{tabular}{|c|c|c|c|c|c|c|c|c|c|c|}
\hline
&0&1&2&3&4&5&6&7&8&9\\
\hline
D$3$ &$-$&$-$&$-$&$-$&$\cdot$&$\cdot$&$\cdot$&$\cdot$&$\cdot$&$\cdot$\\
\hline
D$7$ &$-$&$-$&$-$&$-$&$-$&$-$&$-$&$-$&$\cdot$ &$\cdot$\\
\hline
\end{tabular}
\end{center}
and preserves $8$ supersymmetries as well as a $SO(4) \times SO(2)$
isometry.  Note that separating the D3-branes from the D7-branes in
the 89 direction would explicitly break the SO(2).

Following \cite{afm, Grana, Burrington}, we make the general metric
ansatz
\begin{equation} \label{10dmetric}
ds_{10}^2=h^{-1/2}(y_m)dx_\mu^2 + h^{1/2}(y_m)g_{mn}dy^mdy^n  \,,
\end{equation} 
where $x^\mu$ ($\mu=0,\ldots,3$) are coordinates on the
longitudinal spacetime, and $y^m$ ($m=4,\ldots,9$) are coordinates
transverse to the D3-branes.

The six-dimensional K\"ahler metric transverse to the D3-branes 
is completely fixed by the D7-brane metric 
\be ds_6^2=dz_1d\bz_1+dz_2d\bz_2+e^{\Psi(z_3, \bar z_3)} dz_3d\bz_3 \,,
\label{d7metric} \ee
where the function $e^{\Psi}$ is given by
\begin{align} \label{exppsi}
e^{\Psi(z_3, \bar z_3)} = \tau_2(z_3) \vert 
\eta(\tau) \vert^4 \vert z_3 \vert^{-{N_f}/{6}} \,.
\end{align}
For a review on the properties of the D7-brane metric, see
App.~\ref{appD7}.

We now have to make a choice for the complex structure $\tau$. The
full D3/D7 solution would require the (full) complex structure $\tau$
of the D7-branes as given by Eq.~(\ref{jfu}) in the appendix. However,
for reasons which become apparent in Sec.~\ref{sec25}, we consider
only the weak coupling region of the D7-branes, in which the complex
structure is well-approximated by
\begin{equation} 
 \tau(z)=-i\frac{N_f}{2\pi}\log 
 \left(\frac {z_3} {\rho_L}\right) \,.
\end{equation}
The integration constant $\rho_L$ is given by
\begin{align} 
\rho_L = e^{\frac{2\pi}{g_s N_f}} \,,  
\end{align}
such that $e^{\phi} = g_s$ for $N_f=0$.

This corresponds to focusing in the region close to the D7-branes,
$|z| \ll \rho_L$, where $e^\Psi$ simplifies to
\begin{align}
e^{\Psi(z_3, \bar z_3)} \approx \tau_2(z_3) 
= \frac{1}{g_s}-\frac{N_f}{4\pi}\log z_3\bz_3 \,. \label{logrun}
\end{align}

It has been shown in \cite{Grana} that this ansatz preserves the right
amount of supersymmetry. What remains to be done is to derive an
analytical expression for the warp factor $h(y_m)$ in the metric
ansatz~(\ref{10dmetric}).

\subsection{The warp factor}\label{sec:warpfact}

In order to find an analytical expression for the warp factor, we must
solve the transverse Laplacean
\begin{equation}
\Box_{(6)} h = (\partial_{1}\partial_{\bar 1}+\partial_{2}\partial_{\bar 2}
+e^{-\Psi}\partial_{3}\partial_{\bar 3}) h \,.
\end{equation}
This expression for the warp factor was first examined in \cite{afm},
where it was solved to a first order approximation $z_3=Z_3+\delta z_3$
around an arbitrary but fixed point $Z_3$ away from the D7-branes. 

We next introduce real coordinates on the space transverse to the D3-branes,
\begin{equation}
r^2=|z_1|^2+|z_2|^2,\qquad z_3=\rho e^{i\varphi},
\end{equation} 
in terms of which the transverse metric (\ref{d7metric}) becomes 
\begin{align}
 ds^2_6=d{\vec r}d{\vec r} +
 e^\Psi(d\rho^2+\rho^2d\varphi^2) \,. 
\end{align}
In these coordinates the Laplacean reads 
\begin{equation}
\Box_{(6)} h(r, \rho, \varphi) =
 \left(\fft1{r^3}\partial_rr^3\partial_r+e^{-\Psi(\rho,\varphi)}
 \left(\fft1{\rho}\partial_{\rho}\rho\partial_{\rho}+\fft1{\rho^2}
 \partial_{\varphi}^2\right)\right) h(r, \rho, \varphi) \,.
\end{equation}

The warp factor is, in fact, the Green's function of the Laplace
equation defined by
\begin{align} \label{Poison}
 \Box_6 G(\rho,\varphi,\vec r;\rho', \varphi',\vec
{r'})=N_c e^{-\Psi}\delta^4(\vec r-\vec{r'}
)\delta(\rho-\rho') \frac{1}{\rho}\delta(\varphi-\varphi') \,.
\end{align} 

We solve for the Green's function by performing an expansion in
Fourier modes 
\be 
\delta^4(\vec r-\vec{r'})\delta(\varphi-\varphi')=\frac{1}{(2\pi)^5} 
\int d^4 q e^{i\vec q\cdot (\vec r-\vec r')} \sum_le^{il(\varphi-\varphi')} 
\ee 
and similarly 
\be 
G(\rho,\varphi,\vec r;\rho', \varphi',\vec {r'})=1+Q_{D3} 
\sum_{l} \int d^4 q e^{i\vec 
q \cdot (\vec r-\vec r')}
y_{l, q}(\rho; \rho')e^{il(\varphi-\varphi')} \,,
\ee 
where $Q_{D3}=4\pi g_sN_c l_s^4$ is the D3-brane charge.
Substituting both the expansion of the Green's function and that 
of the delta function into (\ref{Poison}), we find 
\bea
-\frac{l^2}{\rho^2} e^{-\Psi}y_{l,q}(\rho;\rho')+e^{-\Psi} \frac
1\rho\partial_\rho\rho \partial_\rho
y_{l,q}(\rho;\rho')-{q^2}y_{l,q}(\rho;\rho')=\frac{2g_s l_s^4}{(2\pi)^4}
e^{-\Psi}\frac{1}{\rho}\delta(\rho-\rho')  \, \label{green}
\eea 
with $q=|\vec q|$.

Then for $\rho \neq 0$, Eq.~(\ref{green}) leads to the differential equation
\begin{align} \label{de2}
\left( - \frac{\partial^2}{\partial \rho^2} 
-\frac 1{\rho} \frac{\partial}{\partial \rho} + V(\rho) + \frac{l^2}{\rho^2} 
 \right) y_{l, q} (\rho; \rho') = 0 \,,
\end{align}
with the logarithmic potential 
\begin{align}
V(\rho)= \left(\frac{1}{g_s}-\frac{N_f }{2 \pi} \log {\rho} \right) q^2 \,.
\end{align}
By means of the redefinition of the radial coordinate
\begin{align}
 x=\log \rho/\rho_L \,, 
\end{align}
the differential equation (\ref{de2}) turns into 
\begin{align} \label{de3}
( \partial^2_x - l ) y_{l, q} (x) = \lambda x e^{2x} y_{l, q}(x) \,,
\qquad \lambda=\frac{-N_f}{2 \pi} \rho_L^2 q^2  \,,
\end{align}
where we defined $y(x)\equiv y(\rho(x); \rho'=0)$.
We note that in the new coordinate $x$, the near-core region 
($\rho \ll \rho_L$) is located near to $x \rightarrow - \infty$.

There are two independent solutions to the differential equation (\ref{de3}):
they are distinguished by their behavior near the origin and at infinity.
Let us denote by $\tilde y_{l,q}(\rho)$ the solution that is well 
behaved near the origin (if $N_f=0$, this solution reduces to the modified 
Bessel function $I_l(q\rho)$), and by $y_{l,q}(\rho)$ the solution which
is well behaved at infinity (similarly, for $N_f=0$, this solution
becomes the modified Bessel function $K_l(q\rho)$). The explicit 
construction of these two independent solutions will be addressed at length 
in the paragraphs below. The Green's function 
$G(\rho,\dots;\rho',\dots)$ is then obtained by taking the product 
$\tilde y_{l,q}(\rho_<) y_{l,q}(\rho_>)$, where $\rho_<=min(\rho,\rho')$
and $\rho_>=max(\rho,\rho')$. 

Without loss of generality we proceed to construct the Green's function 
$G(r,\rho,\varphi;0)$ 
\begin{align}
G(r, \rho, \varphi; 0)&= 1+  Q_{D3} \int^\infty_0 \frac{dq}{4\pi^3}
q^3 \int_{-1}^1 dt \sqrt{1-t^2} e^{itq |\vec r|}
\sum_{l=-\infty}^{\infty} \tilde y_{l, q}(0) y_{l, q}(\rho)e^{il\varphi} \nonumber \\ 
&= 1+Q_{D3} \int^{\infty}_0
dq \frac{(qr)^2J_1(qr)}{4\pi^2 r^3} 
y_{0, q}(\rho) \,, \label{green2}
\end{align}
in which case we use that $\tilde y_{0,q}(0)$ is the only 
non-vanishing term. This restricts the sum over $l$ to one term
corresponding to $l=0$.

\subsubsection*{Near-core solution of Eq.~(\ref{de3})}

Before deriving the general solution of (\ref{de3}), let us first
consider the solution in the near-core region of the D7-branes. Here it
is convenient to define the coordinate
\begin{align}
\tilde \rho^2 = \lambda x e^{2x} \,.
\end{align}
The corresponding differentials satisfy
\begin{align}
2 \tilde \rho d \tilde\rho=\lambda e^{2x} (1+2x) dx \,.
\end{align}

In the near-core region, close to $x \rightarrow -\infty$, we may
neglect the term $\lambda e^{2x} dx$ on the right hand side of (\ref{de3}) 
and obtain
\begin{align}
d \tilde\rho \approx \tilde \rho dx \,.
\end{align}

In this approximation, Eq.~(\ref{de3}) reduces to the differential 
equation ($y=y_{0, q}$)
\begin{align}
\frac{d}{d\tilde \rho} \tilde \rho  \frac{d}{d\tilde \rho} y(\tilde \rho) =
\tilde \rho y(\tilde \rho) \,,
\end{align}
whose general solution is
\begin{align}
y(\tilde \rho) &= c_1 I_0(\tilde \rho) + c_2 K_0(\tilde \rho) 
\label{solNf0} \,,
\end{align}
where $c_1$ and $c_2$ are constants, and $I_0$ and $K_0$ are the
modified Bessel functions of the first and second kind, respectively.
Since we are interested in solutions which are singular at the
location $\tilde \rho=0$ of the D7-branes and well behaved at infinity, 
we set $c_1=0$, $c_2=
2\pi^2$ and obtain
\begin{align}
y(\tilde \rho)=2\pi^2 K_0(\tilde \rho) \,.
\end{align}

Substituting this solution into (\ref{green2}), we
find the near-core harmonic function
\begin{align} \label{warpnc}
h(r,\rho)&=1+Q_{D3} \int_0^\infty dq q^2 \frac{J_1(q r)}{2 r}
{K_0(\sqrt{\lambda x e^{2x}})} \nonumber \\
&= 1+\frac{Q_{D3}}{(r^2-\rho^2
\frac{N_f}{2\pi} \log\frac{\rho}{\rho_L} )^2}
=  1+\frac{Q_{D3}}{(r^2+\rho^2 e^\Psi)^2}\,,
\end{align}
where we used the identity
\begin{align}
\lambda x e^{2x} = \rho^2 q^2 \left(\frac{1}{g_s}- \frac{N_f}{2\pi} 
\log {\rho} \right) = \rho^2 q^2 e^{\Psi} \,.
\end{align}
For $N_f=0$, we recover the harmonic function of D3-branes
as required. 

The near-core solution can be generalized to the case when the D7-branes are
separated from the D3-branes by a distance $d$ in the overall transverse
plane. Then, the function $e^{\Psi}$ has the form
\begin{equation}
e^{\Psi(\rho, \varphi)}=\frac{1}{g_s}-\fft{N_f}{4\pi}\log(\rho^2-2\rho
d\cos\varphi+d^2) \,.
\end{equation}
The distance $d$ corresponds to the mass $m$ of the $N_f$ flavors.
\subsubsection*{Exact solution of Eq.~(\ref{de3})}

Studying the Schr\"odinger equation of electrons in a logarithmic potential,
 Gesztesy and Pittner (GP) \cite{Gesztesy} found a particular solution
to (\ref{de3}) which is however not the most general one. As shown in
Appendix~\ref{appGP}, the GP solution is not of immediate use for 
the D3/D7 system,
since it reduces to $I_0(\rho q)$ rather than $K_0(\rho q)$ in the
absence of D7-branes. It is therefore not possible to recover the
harmonic function of D3-branes in the limit $N_f=0$. In the following,
we construct a solution for a finite number of flavors, $N_f \neq 0$,
which reduces to the D3-brane solution for $N_f=0$.

Let us first consider the asymptotic behavior of solutions to (\ref{de3}). 
At $x \rightarrow -\infty$, there are two different solutions, 
\begin{align}
y(x) = c \qquad \textmd{and}\qquad y(x)=ax+b \,,
\end{align}
with arbitrary constants $a, b, c$ ($a\neq0$). The GP solution \cite{Gesztesy}
behaves like $y(x)=I_0(0)=1$ at $x \rightarrow -\infty$. We are
however interested in solutions with a singular behavior at $x
\rightarrow -\infty$. In particular, 
the sought after solution should behave as
\begin{align}
y(\rho) = K_0(\rho q) \approx  -\log (\rho q/2)-\gamma  \,.
\end{align}
As discussed in Appendix~\ref{appGP}, this is equivalent to 
choosing the boundary condition $y(x)=ax+b$ with 
\begin{align}
 a=-1 \,, \qquad b=-x_0 - \gamma \,,  \label{bdy}
\end{align}
where $x_0=\log \rho_L q/2$ and $\gamma=-\psi(1)$ is the
Euler-Mascheroni constant. 

Following \cite{Gesztesy}, the solutions with this asymptotic behavior
are given by the series
\begin{align}
y(x)=2\pi^2 \sum_{n=0}^\infty \lambda^n e^{2nx} p_{n} (x)
\,, \label{KV}
\end{align}
where the polynomials $p_n(x)$ are defined recursively by
\begin{align} \label{poly}
\left( 4n^2+4n \frac{d}{dx} + \frac{d^2}{dx^2} \right) p_{n}(x) &=x
p_{n-1}(x) \,, \qquad n=1,2,3,... \,,\nonumber\\
p_{0}(x) &= -x -x_0-\gamma \,.
\end{align}
The recurrence relations are spelled out in Appendix~\ref{appGP}
along with the proof that the series expansion (\ref{KV})
converges uniformly for $x \in (-\infty, 0)$. Moreover, we included another
check that for $N_f=0$ the solution given by (\ref{KV}) 
reduces to the Bessel function $K_0(\rho q)$.
Indeed, if we substitute $y_{0,q}=2\pi^2 K_0(\rho q)$ into (\ref{green2}), we
recover the harmonic function of D3-branes, as expected for $N_f=0$.  This
proves that the series (\ref{KV}) provides the correct analytic solution to the
differential equation (\ref{de3}).

\begin{figure}
\begin{center}
\includegraphics[scale=0.9]{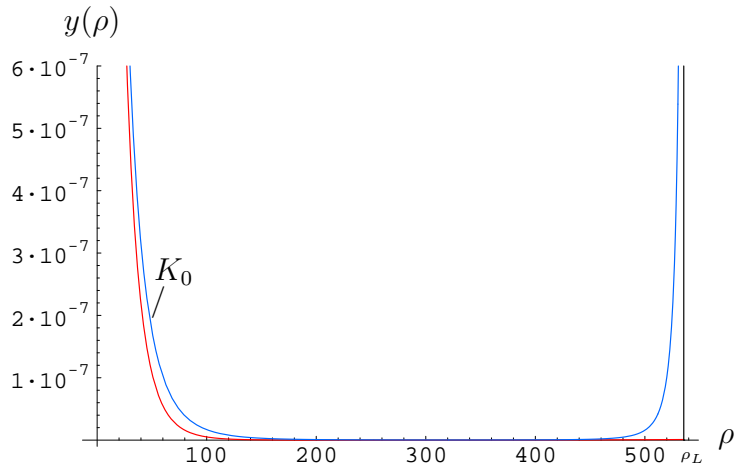} 
\end{center}
\caption{Plot of the near-core solution $y(\rho)=K_0(\sqrt{\lambda x
e^{2x}})$ ($x=\log \rho/\rho_L$) and the exact solution $y(\rho)$
for $N_f=1$, $q=0.1$. The near-core solution breaks down around $\rho_L$.
}
\label{figwarp}
\vspace{-2.7cm} \hspace{12.7cm} $\rho$    \vspace{2.2cm} 

\vspace{-8.2cm} \hspace{4cm} $y(\rho)$    \vspace{7.7cm} 

\vspace{-2.5cm} \hspace{12.2cm} {\tiny $\rho_L$} \vspace{2cm}

\vspace{-4.9cm} \hspace{5.2cm}  {$K_0$} \vspace{4.4cm}

\vspace{-4.5cm} \hspace{5.15cm}  {$/$} \vspace{4cm}
\end{figure}

Upon substituting (\ref{KV}) in (\ref{green2}), we finally obtain the warp
factor
\begin{align} \label{KVsol}
h(r, \rho) &= 1+Q_{D3} \int_0^\infty dq \frac{(qr)^2J_1(qr)}{2 r^3}
  \sum_{n=0}^\infty \lambda^n e^{2nx} p_{n} (x)   \,,
\qquad x=\log (\rho/\rho_L) \,, 
\end{align}
where the polynomials $p_n(x)$  are given by Eq.~(\ref{poly}).
In Fig.~\ref{figwarp} we compare the near-core solution with the
exact solution for $y(\rho)$. In contrast to the exact solution, the
near-core solution is only valid for small $\rho$ and diverges at
$\rho_L=e^{2\pi/N_f}$.

\subsection{The decoupling limit and regime of validity}

We shall now discuss the regime of validity of the supergravity
solution. As mentioned earlier, it is in principle possible to find a
(numerical) solution to the full D3/D7 configuration using the
complete axion-dilaton of the D7-branes. Our background
(\ref{10dmetric}, \ref{d7metric}, \ref{logrun}) approximates this
solution in the regime
\begin{align}
 0 \ll \rho \ll \rho_L \,. 
\end{align}
This regime is bounded from above due to the naked singularity at
$\rho_L$, where the dilaton diverges. The other limit is basically
$\rho=0$, where the space has a curvature singularity due to the
presence of the D7-branes. At this point the string coupling goes to
zero which corresponds to the fact that the dual field theory becomes
free in the infrared.

The decoupling limit is obtained as follows. The massless open string
degrees of freedom correspond to a ${\cal N}=4$ super-Yang-Mills
multiplet (3-3 strings) coupled to $N_f$ bifundamental $\N=2$
hypermultiplets (3-7 strings) localized at the $3+1$ dimensional
intersection. We take a limit in which the 7-7 strings decouple,
leaving a purely four-dimensional theory. This decoupling is achieved
by taking a large $N_c$ limit while keeping the four-dimensional
't~Hooft coupling $\lambda = g^2_{YM} N_c \sim g_s N_c$ and $N_f$
fixed.  This is the usual 't~Hooft limit for the gauge theory
describing the $N_c$ D3-branes. The eight-dimensional 't~Hooft
coupling $\lambda'$ for the $N_f$ D7-branes is $\lambda' = \lambda
(2\pi l_s)^4 N_f/N_c$ which vanishes in the $\alpha' \rightarrow 0$
limit. This turns the $SU(N_f)$ gauge group on the D7-branes into a
global flavor group.

Moreover, we take the limit of large 't~Hooft coupling. Then, the
near-horizon geometry is obtained, as usual, by dropping the ``1'' in
the warp factor $h(r, \rho)$. Writing $g_s e^\Psi= 1- \nu \lambda \log
\rho/\rho_L$ with $\nu=N_f/N_c$, one can easily see that in the strict
$\nu \rightarrow 0$ limit, the near horizon geometry reduces to $AdS_5
\times S^5$ with embedded probe D7-branes~\cite{Karch}. Recall that in
this limit the warp factor (\ref{KVsol}) reduces to the harmonic
function of the D3-branes. In other words, the background includes
corrections to the quenched approximation of first order in $N_f/N_c$.
In the field theory these corrections correspond to Feynman diagrams
with empty quark loops.

The solution seems to be valid for any number of flavors. Neither the
perturbative field theory nor the supergravity solution imposes any
constraint on the number of flavors. Note, however, that a D7-brane
solution, which extends to $\rho \rightarrow \infty$, is known to
exist only for $N_f \leq 12$. It is therefore not clear, whether our
D3/D7 solution can be extended to a full solution in the case of more
than $12$ flavors. It should be safe to use it for $N_f \leq 12$, in
which case the solution can in principle be numerically continued to
infinity. 

\subsection{The warp factor of a fractional D3/D7 brane system}
\label{secnew}

As a bonus for finding the warp factor for the D3/D7 brane system in
flat space, we can actually construct the warp factor of another
system, that of fractional D3/D7 branes on the orbifold
$\RR^{5,1}\times \CC^2/\ZZ_2$, where the $\ZZ_2$ action is a
reflection parity along the $x^4,\dots x^7$ coordinates. The D3-branes
span the $x^0,\dots x^3$ coordinates and the D7-branes are oriented
entirely along the orbifold, spanning $x^0, \dots x^7$. Due to this
fact, the D7-branes also source, beside the axion-dilaton, a twisted
scalar and a twisted 4-form potential. Their charge is half that of
ordinary D7-branes, for which reason they are called fractional D7
branes. The system whose warp factor we will construct next is
composed of $N_f$ fractional D7-branes and $N_c$ fractional D3-branes
\cite{berto}.  The warp factor has been known so far only in terms of
the differential equation 
\bea
(\Box_4+e^{-\phi}\Box_2)H&+&2\kappa^2_{orb}\tau_3 N_c \delta(x^4)\dots
\delta(x^9)\nn\\&+&(2\pi\alpha'g_s)^2\frac{(2N_c-N_f/2)^2}{\rho^2(1-N_fg_s/
(2\pi)\ln(\rho/\epsilon))^3}\delta(x^4)\dots
\delta(x^7)=0\label{fracd3d7} \,, 
\eea 
where $\tau_3$ is the tension
of the fractional D3-brane, $\Box_4$ denotes the four-dimensional flat
space Laplacean along the directions $\{x^4\dots x^7\}=\{\vec r\}$ and
$\Box_2$ is the two-dimensional flat space Laplacean along the
directions $\{x^8, x^9\}\equiv\{\vec\rho\}$. Also, the dilaton is 
\be
e^{-\phi}=1-\frac{N_fg_s}{2\pi}\ln\bigg(\frac{\rho}{\epsilon}\bigg)\,. 
\ee
We recognize the differential operator acting on the warp factor being
the same as the one whose Green's function we have computed previously
\be 
G(\rho,\varphi;\vec r;\rho',\varphi',\vec r')=\frac{C}{(2\pi)^5}
\sum_l\int d^4\vec q e^{i\vec q(\vec r-\vec
r')}e^{il(\varphi-\varphi')} \tilde y_{l,q}(\rho_<)y_{l,q}(\rho_>) \,,
\ee 
where $\vec \rho=\{\rho,\varphi\}$, $\rho_{<}=min(\rho,\rho')$,
$\rho_>= max(\rho,\rho')$, and $\tilde y_{l,q}(\rho_<)$ is the GP
solution, well behaved at the origin, and $y_{l,q}(\rho_>)$ is the
generalization of the solution discussed in App.~B.  Therefore, the
full warp factor of the fractional D3/D7 system is given by 
\bea
H(\vec\rho,\vec r)&=&1-\int d^4\vec r' \int d^2\vec \rho' G(\vec\rho,\vec
r;\vec\rho',\vec r') \bigg[2\kappa^2_{orb}\tau_3 N_c \delta^4(\vec
r')\delta^2(\vec\rho')\nn\\ &+&
(2\pi\alpha'g_s)^2\frac{(2N_c-N_f/2)^2}{|\vec\rho'|^2(1-N_fg_s/
(2\pi)\ln(|\vec\rho'|/\epsilon))^3}\delta^4(\vec r')\bigg] 
\eea 
with
the numerical constant $C=-1$ fixed by requiring that the
discontinuities of Eq.~(\ref{fracd3d7}) at $\rho=\rho'$ match on both
sides.  It is worth saying that the dual gauge theory of this
supergravity background is an $d=4, \N=2$ supersymmetric Yang-Mills
theory, with $SU(N_c)$ gauge group and $N_f$ fundamental
hypermultiplets.  The orbifold action has projected out the adjoint
hypermultiplet of the gauge theory dual to an ordinary D3/D7 system,
leaving a gauge theory with a negative beta function for $N_f<2N_c$.
This means that the fractional D3/D7 supergravity background with a
logarithmically running axion-dilaton is dual to the $\N=2$ gauge theory
at energies {\it above} the scale where the gauge coupling diverges
and where non-perturbative effects become important.

\subsection{The dual $\N=2$ field theory} \label{sec25}

The $\N=2$ field theory dual to the near-horizon geometry of the
D3/D7 configuration in flat space has been studied in 
\cite{Karch, Hong:2003jm}. This theory has $U(N_c)$ gauge group and the field
content of $\N=4$ super Yang-Mills and $N_f$ hypermultiplets in the
fundamental representation. The global symmetry of the theory is
$SO(4) \approx SU(2)_{\Phi} \times SU(2)_{\cal R}$ which consists of
an $SU(2)_\Phi$ global symmetry rotating the scalars in the adjoint
hypermultiplet and an $SU(2)_{\cal R}$ $\N=2$ R-symmetry.  In the
case of overlapping D3 and D7-branes, the fundamentals are massless
and the theory has an additional $U(1)_{\cal R}$ R-symmetry.

We are interested in how far the supergravity background reflects the
perturbative aspects of the gauge theory.  The exact perturbative
$\N=2$ beta function is given by
\begin{align}
\beta_{\N=2} &\equiv \mu \frac{d}{d\mu} \alpha =- \frac{\alpha^2}{2\pi}
\left[ 2 T_G - T_{\rm adj} - N_f T_{\rm fund} \right]
=\beta_{\N=4} + \frac{\alpha^2}{2\pi} N_f = \frac{\alpha^2}{2\pi} N_f \,,
\end{align}
where $\alpha=g^2_{YM}/4\pi$, $T_G=N_c$, $T_{\rm adj}=2T_G$, $T_{\rm
fund}=1$ and $\beta_{\N=4}=0$. In the 't~Hooft limit, the theory is
considered at large $N_c$ with $g_{YM}$ small such that the 't~Hooft
coupling $\lambda=g^2_{YM} N_c$ remains fixed. In this limit a much
more meaningful quantity is the beta function of the 't~Hooft coupling
\begin{align}
\beta^\lambda_{\N=2} \equiv N_c \beta_{\N=2} = \frac1{2\pi}
\left(\frac{\lambda}{4\pi} \right)^2 \frac{N_f}{N_c} \,.
\end{align} 
We see that if $N_f/N_c$ is kept fixed, then the field theory is
neither conformal nor asymptotically free. The theory is however
conformal in the strict $N_c \rightarrow \infty$ limit \cite{Karch}.

A positive beta function leads to a running gauge coupling
$g_{YM}$ which is given by
\begin{align} \label{ymcoupling}
\alpha(Q^2) = \frac{1}{\frac{\beta_0}{4\pi} \log
 \frac{\Lambda^2_L}{Q^2}} \qquad{\textmd{with}} \qquad \Lambda^2_L =
 \mu^2 \exp\left( \frac{4\pi}{\beta_0 \alpha(\mu^2)} \right)\,,
\end{align}
$Q^2$ the energy scale, $\mu^2$ a reference scale and $\beta_0=N_f$.
$\Lambda_L$ is the scale of the Landau pole at which the gauge
coupling becomes divergent.\footnote{In the $\N=2$ theory of the
orbifolded D3/D7 intersection considered in Sec.~\ref{secnew}, adjoint
hypermultiplets are projected out, and the beta function $\beta^\lambda \sim
-(2-N_f/N_c)$ is negative, if $N_f < 2N_c$.}

This can be compared with the action for a probe D3-brane placed in the
D3/D7 supergravity background (\ref{10dmetric}, \ref{d7metric}, \ref{logrun}),
\begin{align}
S_{D3}&= -T_{D3}\int d^4 \sigma e^{-\Phi} \sqrt{-\det (g_{ab}
+ {\cal F}_{ab}) } + T_{D3} \int C^4
+ C^0 {\cal F} \wedge {\cal F} \nonumber\\
&\approx T_{D3} \int d^4 \sigma (2\pi \alpha')^2\left[ 
- \frac{1}{4} e^{-\Phi} F_{ab}F^{ab} + \chi F_{ab} \tilde F^{ab} 
\right] + ... \,,   \label{DBI}
\end{align}
where ${\cal F}_{ab} =(2\pi \alpha')F_{ab}$. Here we expanded to
second order in the field strength and kept only terms relevant for
the field theory. The axion $C^0=\chi$ and the dilaton are given by
the complex structure (\ref{tauperturb}). They are
\begin{align} \label{dilaton}
\chi=\frac{N_f}{2\pi} \varphi \,, \qquad
e^{-\Phi}
= {\frac{N_f}{4\pi} \log \frac{\rho^2_L}{\rho^2}} \,. 
\end{align} 

The D3-brane action relates the dilaton and the axion to the gauge
coupling $g_{YM}^2=4\pi e^{\Phi}$ and theta angle
$\theta_{YM}=2\pi\chi$. Upon identifying also $\alpha(Q^2) =
g^2_{YM}(Q^2)/4\pi$, $\alpha(\mu^2)=g^2_{YM}(\mu^2)/4\pi=g_s$,
$Q^2=\rho^2/(2\pi \alpha'{}^2)$, $\Lambda_L=\rho_L/(2\pi
\alpha'{}^2)$, the running of the gauge coupling~(\ref{ymcoupling})
follows from the logarithmic behavior of the dilaton, and the Landau
pole $(g_{YM}\to \infty)$ is related to the dilaton divergence at
\begin{align} 
\rho_L= e^{\frac{2\pi}{g_s N_f}} = e^{\frac{2\pi N_c}{\lambda N_f}}  \,.  
\end{align} 
This shows that both sides of the duality show the same pathology: The
perturbative field theory becomes strongly coupled at the Landau pole
$\Lambda_L$, while the supergravity solution breaks down at some
distance $\rho_L$.\footnote{In the conformal limit $N_f/N_c
\rightarrow 0$, the location $\rho_L$ is formally shifted to infinity
and the near-horizon geometry of the solution reduces to $AdS_5 \times
S^5$ with embedded probe D7-branes.}

Furthermore, the $U(1)_{\cal R}$ chiral anomaly in the field theory is
reflected by a nontrivial axion profile in the supergravity background.
The instanton term in (\ref{DBI}) correctly reproduces the Yang-Mills
theta angle $\theta_{YM}=2\pi \chi={N_f}\varphi$. This corresponds to
the breaking of the classical $U(1)_{\cal R}$ symmetry to $\ZZ_{2N_f}$ at the quantum level.

This shows that the solution (\ref{10dmetric}, \ref{d7metric}, \ref{logrun}) of the D3/D7 system
perfectly reflects the perturbative aspects of the dual $\N=2$ field
theory.

\subsection{Logarithmic tadpoles and one-loop vacuum amplitudes}

Since D7-branes are codimension-two branes, we expect uncanceled
tadpoles in the string background.  Tadpole divergences usually
correspond to gauge anomalies and indicate an inconsistency in the
theory. However, as it was found first in \cite{Leigh:1998hj},
logarithmic tadpoles do not correspond to gauge anomalies, but reflect
the fact that the dual gauge theory is not conformal invariant. In
fact, as we will show now following arguments given in
\cite{DiVecchia:2005vm}, such tadpoles provide the correct one-loop
running of the gauge coupling.

This can be seen by considering the one-loop annulus diagram of an
open string stretching between the stack of D7-branes and a D3-brane
dressed with an external background $SU(N_c)$ gauge field. In the
field theory limit this vacuum amplitude reduces to the one-loop
diagrams determining the running of the gauge coupling
\cite{DiVecchia:2005vm}. When expanded to quadratic order in the gauge
field, the amplitude reads:\footnote{This one-loop open string
amplitude is obtained by summing up the one-loop amplitudes of the
twisted and untwisted sector of the orbifolded D3/D7 configuration
which have been computed in \cite{DiVecchia:2005vm}; it is identical
to $2Z_{e;37}= (Z_{e;37}+ Z_{h;37})+(Z_{e;37}-Z_{h;37})$ given by
Eq.~(75) therein. Note also that the amplitude is different from the one
corresponding to a string stretching between the stack of D3-branes
and the stack of D7-branes. The latter vanishes as expected for a BPS
Dp-D(p+4) brane configuration.}
\begin{align} \label{annulus}
Z_{37} = \frac{N_f}{(4\pi)^2} \int d^4 x \left[ -\frac{1}{4} \left(F^a_{\mu\nu}
F^{a\mu\nu} - i F^a_{\mu\nu} \tilde F^{a\mu\nu} \right) \right]
\int_{1/(\alpha'\Lambda^2)}^\infty \frac{d\tau}{\tau} e^{-\frac{\rho^2 \tau}
{2\pi\alpha'}} \,.
\end{align}
 The integral over $\tau$ is logarithmically divergent for small
$\tau$ and has been regularized by an ultraviolet cut-off~$\Lambda$.
The parameter $\rho$ is the distance between the dressed D3-brane and
the D7-branes. Threshold corrections to the amplitude (\ref{annulus}),
coming from massive string states, are absent in the $\alpha'
\rightarrow 0$ limit.

From the annulus amplitude, one deduces the gauge coupling
\begin{align}
\frac{1}{g_{YM}^2} =  \frac{N_f}{(4\pi)^2} \int_{1/(\alpha'\Lambda^2)}^\infty
\frac{d\tau}{\tau} e^{-\frac{\rho^2 \tau} {2\pi\alpha'}} =
 \frac{N_f}{(4\pi)^2} \log \frac{\Lambda^2}{Q^2} \,,
\end{align}
where we defined the energy scale $Q^2=\rho^2/(2\pi \alpha'{}^2)$.
The previous relation agrees with Eq.~(\ref{ymcoupling}) and the 
dilaton behavior described by (\ref{dilaton}). This shows that tadpoles 
provide the correct one-loop running of the gauge coupling.

Furthermore, from the instanton term in (\ref{annulus}), we may also
extract the $\theta_{\rm YM}$ angle,
\begin{align}
 \theta_{\rm YM} = N_f \varphi \,,
\end{align}
where $\varphi$ is the phase of the complex coordinate $z=\rho
e^{i\varphi}$ transverse to the D7-branes. As mentioned above, in the
supergravity background the chiral anomaly is encoded in the axion
$\chi$.

Under open/closed string duality, one can consider the above one-loop
amplitude also as a tree-level closed string amplitude which encodes
information about the supergravity solution.  The absence of threshold
corrections, i.e.\ massive open string state contributions to the
gauge coupling, guarantees that open massless string states are
precisely mapped into massless closed string states. The supergravity
solution constructed in this paper is therefore sufficient to describe
the perturbative field theory.  Even though the dual string background
has logarithmic divergences, it is still consistent as far as the
description of the embedded non-conformal field theory is concerned.

Finally, it should be clear that in the conformal limit $N_f/N_c
\rightarrow 0$, considered by Karch and Katz~\cite{Karch}, the gauge
coupling does not run, and by the above arguments, tadpoles are
absent. Certainly, D7-branes which do not backreact on the geometry do
not emit flux and there is no net charge.

\section{Regge trajectories in the D3/D7 theory} \label{sec3}
\setcounter{equation}{0}\setcounter{figure}{0}\setcounter{table}{0}

As an application of our supergravity solution, we now discuss the
general structure of Regge trajectories in the ${\cal N}=2$ gauge
theory of the D3/D7 system by means of a semi-classical string
computation.  This will provide interesting information about the
behavior of the Regge trajectories in dependence of the number of
flavors. We shall compare our results with a similar analysis
performed in the probe approximation in \cite{Kruczenski:2003be}.

\subsection{Spinning strings in the D3/D7 background}

\begin{figure}
\begin{center}
\includegraphics[scale=0.8]{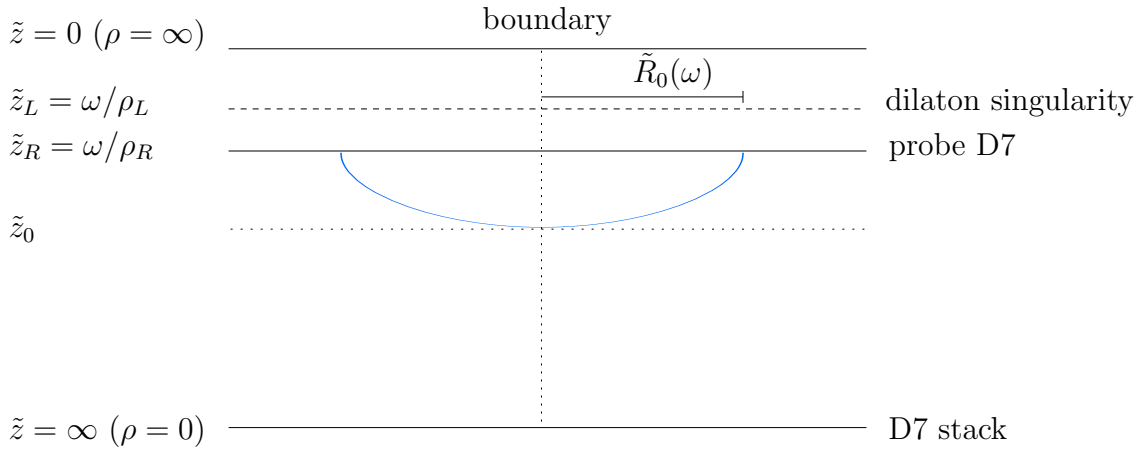} 
\end{center}
\caption{Example of a string profile $\tilde z(\tilde R)$.} \label{figbc}
\vspace{-1.7cm} \hspace{0.7cm} $\tilde z=\infty$ $(\rho=0)$ 
\hspace{8.8cm} D7 stack  \vspace{1.2cm} 

\vspace{-4.4cm} \hspace{0.7cm} $\tilde z_{0}$ \vspace{3.8cm}

\vspace{-5.4cm} \hspace{0.7cm} $\tilde z_R=\omega/\rho_R$ 
\hspace{9.5cm} probe D7 \vspace{4.9cm}

\vspace{-6.0cm} \hspace{0.7cm} $\tilde z_L=\omega/\rho_L$ 
\hspace{9.5cm} dilaton singularity\vspace{5.5cm}

\vspace{-6.9cm} \hspace{0.7cm} $\tilde z=0$ $(\rho=\infty)$ \vspace{6.4cm}

\vspace{-7.1cm} \hspace{7cm} boundary \vspace{6.6cm}

\vspace{-6.4cm} \hspace{9.cm}  $\tilde R_0(\omega)$ \vspace{5.9cm}
\end{figure}

Following \cite{Gubser, Kruczenski:2003be, Kruczenski:2004me}, we
consider an open string rotating in the near-horizon limit of the
D3/D7 background. We parameterize the four-dimensional spacetime part
of our background (\ref{10dmetric}, \ref{d7metric}, \ref{logrun}) as
\begin{align}
dx^\mu dx_\mu &= -dt^2 + dR^2 +R^2 d\varphi^2 +dz^2 \,, 
\end{align}
where $R$ and $\varphi$ are the coordinates of the plane of rotation. The
string has length $2R_0$ and stretches from $-R_0$ to $+R_0$ along the
$R$ direction. The end points of the string are attached to an
additional probe D7-brane located a distance $\rho_R$ away from the
stack of D7-branes. Recall that we have to choose $\rho_R$ in the
regime $0 \ll \rho_R \ll \rho_L$, i.e.\ far away from the singularity
at the location of the stack of D7-branes and from the dilaton
divergence at $\rho_L$.  An example of a spinning string is shown in
Fig.~\ref{figbc}.

In the field theory the set-up corresponds to $N_f$ massless flavors
plus an additional massive flavor whose mass is proportional to
$\rho_R$. In this theory we consider a meson, which consists of
massive flavors, in the presence of $N_f$ massless flavors. We assume
a large spin for the meson which allows a classical treatment of the
dual spinning string. Recall that meson operators with large spin have
small anomalous dimensions and quantum corrections are
negligible~\cite{Gubser}.

An appropriate ansatz for a string rotating with constant angular
velocity $\omega$ is \be t=\tau, \quad \varphi =\omega\tau, \quad
R=R(\sigma), \quad r=r(\sigma), \quad \rho=\rho(\sigma)\,, \ee with
world sheet coordinates $\sigma$ and $\tau$.  With this ansatz the
classical Nambu-Goto action takes the form 
\begin{align} 
{\cal L} = -T_s
\sqrt{(1-\omega^2 R^2)(h^{-1} R'{}^2 +r'{}^2+(1-\frac{N_f}{2\pi}\log
\rho )\rho'{}^2)} \,.  
\end{align} 
It is convenient to use the rescaled coordinates
\begin{align}
\tilde R = \omega R \,,\quad \tilde r=\frac{r}{\omega}
\,,\quad \tilde \rho=\frac{\rho}{\omega} \,.
\end{align}
In these coordinates, the energy and the angular momentum of
the spinning string are given by
\begin{align}
E&=\int d\sigma \left(\omega \frac{\partial {\cal L}}{\partial \omega}
- {\cal L} \right) = \int d\sigma \frac{\omega}{\cal E} \sqrt{\tilde h^{-1} 
\tilde R'(\sigma){}^2
  + \tilde r'(\sigma)^2 + e^\psi \tilde \rho'(\sigma)^2} \,,
\label{energy}\\
J&=\int d\sigma \frac{\partial {\cal L}}{\partial \omega}= \int
d\sigma \frac{\tilde R^2}{\cal E} \sqrt{\tilde h^{-1} \tilde
R'(\sigma)^2 + \tilde r'(\sigma)^2 + e^\psi \tilde 
\rho'(\sigma)^2} \,, \label{spin}
\end{align}
with ${\cal E}= \sqrt{1- \tilde R^2}$, $T_s =1$ and 
\begin{align}
\tilde h(\tilde r,\tilde \rho)=\frac{Q_{D3}}{(\tilde r^2+\tilde \rho^2 
e^{\Psi(\tilde \rho)})^2}\,.
\end{align}

In the gauge $\tilde R=\sigma$, we find the following equations of
motion for $\tilde r( \tilde R)$ and $\tilde \rho( \tilde R)$: 
\begin{align}
 \frac{d}{d \tilde R}\left(\frac{{\cal E}^2}{\cal L}\partial_{\tilde
R} \tilde r\right)&= \frac{{\cal E}^2}{2{\cal L}}\partial_{\tilde r}
\tilde h^{-1}, \label{eomstring1} \\ \frac{d}{d \tilde
R}\left(\frac{{\cal E}^2}{{\cal L}}e^\Psi \partial_{\tilde R} \tilde
\rho\right)&= \frac{{\cal E}^2}{2{\cal L}}\left( \partial_{\tilde
\rho} \tilde h^{-1}+ (\partial_{\tilde R} \tilde \rho)^2
\partial_{\tilde \rho} e^{\Psi}\right) \,. \label{eomstring2}
\end{align} 
The warp factor $\tilde h(\tilde r, \tilde \rho)$ satisfies
$\partial_{\tilde r} \tilde h^{-1}=0$ for $\tilde r=0$ which shows
that $\tilde r \equiv 0$ is a solution of the equation of
motion~(\ref{eomstring1}). Writing Eq.~(\ref{eomstring2}) in
coordinates $\tilde z=1/\tilde \rho$, we find
\begin{align}
\frac{d}{d \tilde R}\left(\frac{{\cal E}^2}{{\cal L}} e^\Psi
\partial_{\tilde R} \tilde z^{-1} \right)= - \frac{{\cal
E}^2}{2{\cal L}} \tilde z^2 \left( \partial_{\tilde z} \tilde
h^{-1}+ (\partial_{\tilde R} \tilde z^{-1})^2 \partial_{\tilde z}
e^{\Psi}\right) \,. \label{eomstring2a}
\end{align}

This is a nonlinear differential equation of second order which
requires two boundary conditions. We impose the usual open string
boundary conditions
\begin{align}
\left.\frac{\partial {\cal L}}{\partial \tilde R'} \delta \tilde 
R \right\vert_{\sigma=0,\pi}
=
\left.\frac{\partial {\cal L}}{\partial \tilde z'} \delta \tilde 
z \right\vert_{\sigma=0,\pi}=0\,
\end{align}
for a string ending on a probe D7-brane at $\tilde z=\tilde
z_R=const$. Due to the Neumann boundary condition in the $\tilde R$
direction and the Dirichlet boundary condition in the $\tilde z$
direction, $\delta \tilde R\vert_{\sigma=0,\pi}$ is arbitrary, whereas
$\delta \tilde z \vert_{\sigma=0,\pi}=0$. The remaining condition
$\partial {\cal L}/{\partial \tilde R'} \vert_{\sigma=0, \pi}=0$ is
satisfied, if \mbox{$\tilde R'\vert_{\sigma=0,\pi}=0$}.  Using the
gauge {$\tilde z=\sigma$}, we see that this corresponds to ${\partial
\tilde z}/{\partial \tilde R}\vert_{\tilde R=\pm \tilde R_0}
\rightarrow \infty$. This means that the string ends orthogonally on
$\tilde z_R$.

We expect the solutions to be symmetric around $\tilde R=0$, where
they have their only maximum.  We thus impose the additional boundary
condition $\tilde z'(0)=0$.  Taking into account the orthogonal ending
of the string on a constant value of~$\tilde z$, we set $\tilde z(\pm
\tilde R_0)=\tilde z_R = const.$ or, equivalently, $\tilde z(0)=\tilde
z_0 =const$. 

\subsection{Dependence of Regge trajectories on the number of flavors}
\label{sec32}

\begin{figure}
\vspace{1.cm}
\begin{center}
\includegraphics[scale=0.9]{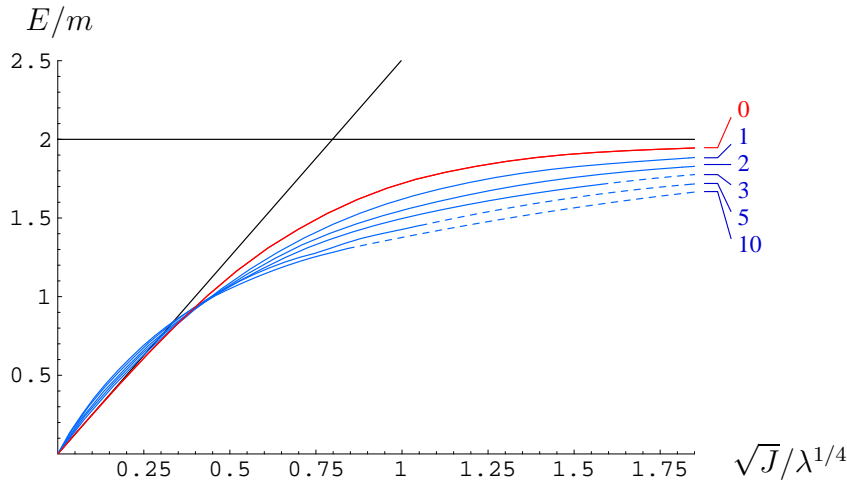} 
\end{center}
\caption{Chew-Frautschi plot for $N_f=0,1,2,3,5, 10$ additional
massless flavors. The straight line represents the $N_f=0$ 
trajectory for small spin values. All graphs approach the
horizontal line $E=2m$.}\label{figregge}
\vspace{-3cm} \hspace{12.4cm} $\sqrt{J}/\lambda^{1/4}$    \vspace{2.5cm} 

\vspace{-8.9cm} \hspace{3cm} $E/m$    \vspace{8.4cm} 

\vspace{-7.9cm} \hspace{12.9cm}     \vspace{7.4cm}

\vspace{-6.9cm} \hspace{12.9cm}     \vspace{6.4cm}

\vspace{-5.8cm} \hspace{12.9cm}    \vspace{5.3cm}
\end{figure}

The Regge trajectories $E(J; N_f)$ can now be obtained as follows.  We
first solve the equation of motion (\ref{eomstring2a}) for the string
profile $\tilde z(\tilde R)$. We then substitute the profile into the
expressions (\ref{energy}) and (\ref{spin}) for the energy $E(\omega)$
and the angular momentum $J(\omega)$ of the spinning string. This
yields a Regge trajectory as a curve in the $\sqrt{J},E$ plane
parameterized by the angular velocity $\omega$.

We begin by fixing the mass $m$ of the single flavor corresponding to
a 3-7 string stretching from the D3-branes at $\rho=0$ ($\tilde z
\rightarrow \infty$) to the probe D7-brane at $\tilde z_R
=\omega/\rho_R$:
\begin{align}
m=\int_{\varepsilon}^{\rho_R} \sqrt{g_{00} g_{\rho\rho}} d\rho =
\int_{\varepsilon}^{\rho_R} e^{\Psi/2} d\rho
=\int_{\varepsilon}^{\rho_R} (\frac 1 {g_s} -\frac{N_f}{2\pi} \log \rho)^{1/2}
d\rho \,,
\end{align}
where $\varepsilon \ll 1$ is a (IR) cut-off which is needed to
regularize the singularity at $\rho=0$. We set $m=1$ and solve this
equation numerically for $\rho_R(m, N_f)$.  In this way we obtain the
location of the probe D7-brane $\tilde z_R(m, N_f)=\omega/\rho_R(m,
N_f)$. The set of Regge trajectories in Fig.~\ref{figregge} is obtained
by varying $\omega$ for a given number of massless flavors $N_f$, then
changing $N_f$ while all the time keeping the mass $m$ fixed.

We now determine the string profile $\tilde z_R$ for given $N_f$ and
$\omega$. To this end, we integrate the equations of motion
(\ref{eomstring2a}) from $-\tilde R_0$ to $+\tilde R_0$. In the
shooting technique, we set $\tilde z(0)=\tilde z_0 =const$, $\tilde
z'(0)=0$ such that $\tilde z(\pm \tilde R_0)=\tilde z_R$. This yields
the string length $R_0=\tilde R_0/\omega$ as the location at which
$\tilde z'(\tilde R_0) \rightarrow \infty$.  A typical profile has
been shown in Fig.~\ref{figbc}.\footnote{In general, one obtains a
series of solutions labeled by the number of nodes of the string
\cite{Kruczenski:2003be}. We restrict to string solutions with no
nodes which are believed to be the most stable ones.}

Substituting the profiles $\tilde r(\tilde R) \equiv 0$ and $\tilde
\rho(\tilde R)^{-1}=\tilde z(\tilde R)$ into Eqns.~(\ref{energy}) and
(\ref{spin}), we determine the energy $E(\omega)$ and the angular
momentum $J(\omega)$ in dependence of the angular velocity
$\omega$. Repeating the procedure for different $\omega$, we obtain a 
parametric plot for the Regge trajectory $E(\sqrt J; N_f)$ associated
with a certain number of flavors $N_f$.

In the computation we keep $g_s \sim \lambda/N_c$ small, but fixed such
that the background depends on the small but finite parameter $\nu=N_f/N_c$.

Our results are presented in Fig.~\ref{figregge} which shows the Regge
trajectories for different numbers of massless flavors
$N_f$.\footnote{For each graph there exists a spin value $J$ at which
the location $\tilde z_R$ of the probe D7-brane becomes of the order
of $\tilde z_L$. Here the solution (\ref{10dmetric}, \ref{d7metric},
\ref{logrun}) breaks down.  The dashed part of the graphs is an
interpolation to higher spin values.}  The energy $E$ and the spin $J$
of the mesons are given in units of the quark mass $m$ and the square
root of the 't~Hooft coupling $\lambda$, respectively.  As a first
check of our numerics, we note that the graph for $N_f=0$ coincides
with that found in the probe limit
\cite{Kruczenski:2003be}. Disregarding the behavior at small spin
values, the graph for $N_f=0$ runs above the graphs for $N_f \neq 0$.

\begin{figure}
\vspace{1.cm}
\begin{center}
\includegraphics[scale=0.8]{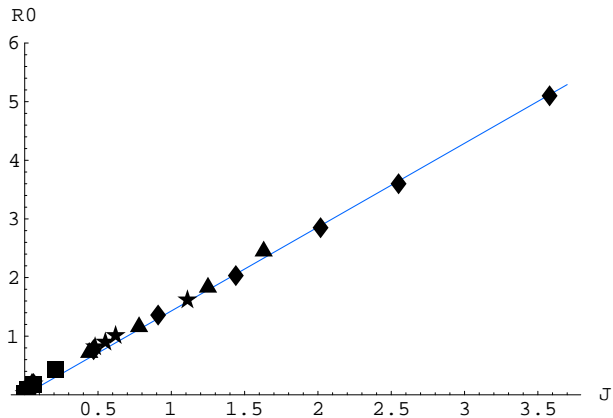} 
\end{center}
\caption{String length $R_0(J)$ for different $N_f$.}\label{figlength}
\end{figure}
The semiclassical approximation is valid as long as the quantum
effects are suppressed by $T_s/S_{cls} \ll 1$ where
$S_{cls}=E/\omega-J$ is the value of the action evaluated on the
classical solution.  One can check that while the linear behavior of
the Regge trajectories lies well outside the region of validity of the
semiclassical approximation,\footnote{The spinning string
characterized by small $E,J$ is not macroscopic and probes a small
region of the warped geometry which, being small, appears almost flat
giving rise to linear Regge trajectories. Quantum effects however will
largely correct this apparent confining behavior.} the region where
the quantum effects are suppressed is still sensitive to the number of flavors.

Let us first consider the general behavior of the Regge
trajectories. Quite generically, the trajectories are linear in the
small spin region and asymp\-tote to the rest energy $E=2m$ in the large
spin limit. This can be understood from the behavior of the string
length as a function of the spin. Fig.~\ref{figlength} shows that the
data fits a linear relation between the string length $R_0$ and the
spin $J$. Note that this plot contains data from graphs for different
$N_f$. Since the linear relation is the same for all $N_f$, we find
that the string length $R_0$ does not depend on the number of massless
flavors.

At small spin values the length of the string is much smaller than the
scale of the space, $R_0 \ll Q_{D3}$, and the string is effectively
rotating in flat space leading to a linear Regge behavior. At large
spin the string is larger than the size of the space, $R_0 \gg
Q_{D3}$. Here the string rotates very slowly and the energy is that of
particles moving in a Coulomb potential \cite{Kruczenski:2003be}.
Moreover, since the string length grows linearly with spin, also the
binding energy of the quark-antiquark pair vanishes at large spin
values. The Regge trajectories thus asymptote to the rest energy
$E=2m$ in the limit $\omega \rightarrow 0$.

Let us now compare trajectories with different numbers of flavors.  We
observe that for a meson with fixed (large) spin, the energy is the
lower the larger the number of flavors in the theory, i.e.~$E_1 > E_2$
for $N_f^1 < N_f^2$ and $J_1=J_2$. In other words, the graph for
smaller $N_f$ is above the graph for higher $N_f$ in the $\sqrt{J},E$
plane. 
\begin{figure}
\begin{center}
\includegraphics[scale=0.8]{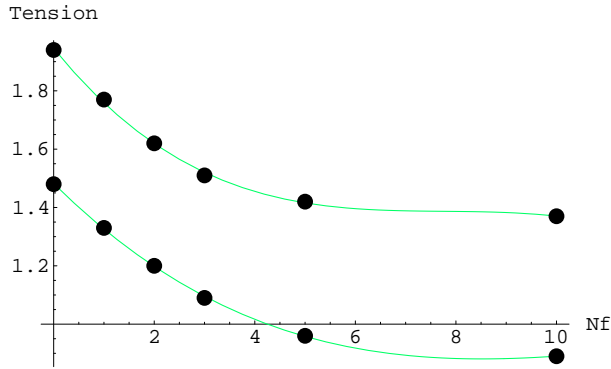} 
\end{center}
\caption{String tension in dependence of the number of flavors $N_f$ for
$J=0.4^2$ (upper graph) and $J=0.6^2$ (lower graph).}\label{figtension}
\end{figure}

To understand this behavior, it is illuminating to consider the string
tension for a fixed spin value as a function of $N_f$. The square root
of the tension is proportional to the derivative of $E(\sqrt
J)$. Fig.~\ref{figtension} shows that the string tension is a
monotonically decreasing function of $N_f$. Since other quantities
like the string length are unaffected by an increasing number of
flavors, the energy of a meson is smaller in a theory with more
flavors. This effect vanishes however at very large values of the
spin.  

The behavior of the string tension can also be understood in the dual
field theory. The rotating string can be considered as a spin chain of
gluons with quarks at the end of the string. In between the quark and
the antiquark of the meson there are virtual quark-antiquark pairs 
which interact with the gluon chain. The polarization of these virtual
quark-antiquark pairs effectively diminishes the color force between
the quarks of the meson. This screening effect should have an
influence on the string tension. We expect that a large number of
flavors in a field theory leads to a small string tension. At least
heuristically, this explains the fact that a meson with a certain spin
has less energy than the same meson in a theory with a larger number
of flavors.  

Even though the background we are discussing is not confining, 
a similar effect is observed in lattice QCD with dynamical light quarks, 
where the interaction potential between two heavy quarks
\be
V(r)=\left(-\frac{\alpha}r+\sigma r\right)\frac{1-e^{-\mu r}}{\mu r}
\ee 
was a better fit to the data, rather than the unscreened potential 
\cite{Born}. For non-vanishing inverse screening length $\mu$, the 
effective flux tube tension between the heavy quarks is diminished 
due to the screening of the color charge, with $\mu$ increasing
with the number of light quarks.

\section{Conclusions} \label{sec4}
\setcounter{equation}{0}\setcounter{figure}{0}\setcounter{table}{0}

To summarize our results, we have completed the type IIB supergravity
solution describing the fully localized D3/D7 brane configuration by
providing an analytic expression for the warp factor.  The background
is given by the metric
\begin{equation} \label{solution}
ds_{10}^2=h^{-1/2}(r,\rho) \, dx_{\mu}^2 
+ h^{1/2}(r,\rho) \left(dr^2+r^2d\Omega^2_3 
+ e^{\Psi(\rho)} (d\rho^2+\rho^2 d\varphi) \right) \,
\end{equation}
for which we assumed a logarithmically running axion-dilaton:
\begin{align} 
\chi(\varphi)=\frac{N_f}{2\pi} \varphi \,, \qquad
e^{-\phi(\rho)}=e^{\Psi(\rho)} = -\frac{N_f}{2\pi}\log \frac{\rho}{\rho_L} \,,
\qquad \rho_L=e^{\frac{2\pi}{g_s N_f}} \,.
\end{align}
Upon taking the decoupling limit,
this is the supergravity dual of ${\cal N}=4$ $SU(N_c)$ super
Yang-Mills gauge theory with $N_f$ ${\cal N}=2$ fundamental
hypermultiplets.

The warp factor $h(r,\rho)$ is given in terms of a convergent series:
\begin{align}
h(r, \rho) &= 1+Q_{D3} \int_0^\infty dq \frac{(qr)^2J_1(qr)}{2 r^3}
  \sum_{n=0}^\infty \lambda^n e^{2nx} p_{n} (x)   \,,
\qquad x=\log (\rho/\rho_L) \,, 
\end{align}
where $\lambda=\frac{-N_f}{2 \pi} \rho_L^2 q^2$, and
the polynomials $p_n(x)$ ($n=0,1,2,...$) are defined recursively by
\begin{align}
\left( 4n^2+4n \frac{d}{dx} + \frac{d^2}{dx^2} \right) p_{n}
&=x p_{n-1} \,, \qquad p_{0}(x) = -x -\log (\rho_L q/2)-\gamma  \,.
\end{align}
In the immediate vicinity of the D7-branes, $x \rightarrow -\infty$ ($\rho
\rightarrow 0$), the warp factor can be approximated by the
simple expression
\begin{align}
h(r,\rho)=1+\frac{Q_{D3}}{(r^2+\rho^2 e^{\Psi(\rho)})^2} \,.
\end{align}

We have related the pathology of the supergravity background to the 
Landau pole of the gauge theory which has a positive beta function. We
have also shown that the existence of a logarithmic tadpole in the one-loop
open string amplitude between the D3 and D7-branes is related to the
same one-loop running of the gauge coupling.

In the second part of this paper, we have considered classical
spinning string configurations in the backreacted D3/D7 geometry and
analyzed the dependence of the associated meson Regge trajectories on
the number of flavors. From the Chew-Frautschi plots $(E,\sqrt{J})$ we
have found that for the same spin, the energy of a meson is smaller in
a gauge theory with more flavors.  We interpreted this in terms of a
screening effect of the color charge due to virtual $q\bar q $ pairs.
Since the background is non-confining, in the IR, where the theory
becomes conformal, the tension of the quark-antiquark flux tube
vanishes at large spin $J$, and the meson configuration energy approaches
the rest mass of its constituents.

We have seen that taking into account the backreaction of the flavor
branes opens up the possibility of going beyond the quenched
approximation via the AdS/CFT correspondence. An immediate goal would
be the study of confining backgrounds, including the backreaction of
the flavor branes.  Also, it is well-known that $SU(N)$ gauge theories
undergo a chiral phase transition \cite{Appelquist:1996dq} as the
number of flavors is varied, even at zero temperature. It would be
interesting if these phase transitions could be studied from the
perspective of the dual supergravity background.

\setcounter{equation}{0}\setcounter{figure}{0}\setcounter{table}{0}

\section*{Acknowledgments}

We would like to thank L.~A.~Pando~Zayas for participation at an early
stage of this project. Moreover, we are grateful to A.~Fayyazuddin,
N.~Arkani-Hamed, Z.~Guralnik, M.~Lublinsky, L.~Motl,
C.~N\'{u}$\rm\tilde n$ez for many helpful discussions related to this
work.

The research of I.K. is supported by a fellowship within the
Postdoc-Programme of the German Academic Exchange Service (DAAD),
grant D/04/23739.


\appendix
\section{Brief review of the D7-branes geometry} \label{appD7}
\setcounter{equation}{0}\setcounter{figure}{0}\setcounter{table}{0}

For completeness, we review here the geometry of a stack of D7-branes 
which has first been studied in~\cite{Greene:1989ya} in the context of 
cosmic string solutions. 
The standard metric ansatz for the type IIB supergravity solution 
corresponding to a stack of $N_f$ coincident D7-branes is 
\be ds_{10}^2=ds_8^2+e^{\Psi(z, \bar z)} dzd\bar z \,, \ee
where $ds^2_8$ is an eight-dimensional flat metric and $z$ is a
complex coordinate parameterizing the space transverse to the
D7-branes. D7-branes couple to the axion-dilaton field
$\tau=\chi+ie^{-\phi}$.

The complex structure $\tau$ and the function $\Psi$ have
to satisfy the equations of motion~\cite{Greene:1989ya}
\begin{align}
\partial \overline{\partial} \tau + \frac{2\partial\tau
\overline{\partial} \tau}{\bar \tau-\tau} =N_f \delta(z)\,,
\label{eqntau}\\ \partial \overline{\partial} \Psi = \partial
\overline{\partial} \,\log {\,\rm Im\,}(\tau) \,. \label{eqnepsi}
\end{align}

The first equation requires $\tau$ to be holomorphic or
anti-holomorphic. In order to have a finite string coupling, ${\rm
Im}\, (\tau)$ must be nonzero for all values of $|z|$. The basic
solution for $\tau$ can be written in terms of the modular invariant
$j$-function~\cite{Greene:1989ya}. A stack of $N_f$ D7-branes
corresponds to a pole of order $N_f$ in the $j$-plane which can
locally be parameterized by
\begin{align} \label{jfu}
  j(\tau)=\frac{1}{(z/\rho_L)^{N_f}} \,.
\end{align}
For small $|z|$, $j(\tau)$ is large and can be approximated by
$\exp(-2\pi i \tau)$. Substituting $j \sim \exp(-2\pi i \tau)$ in
Eq.~(\ref{jfu}) and solving for $\tau$, we get the weak-coupling
approximation\footnote{This is the complex structure which one would
naively expect from the general behavior of the dilaton of a D$p$-brane
($p < 7$), $e^{-\phi}=(1+A/\rho^{7-p})^{(p-3)/4}$.}
\begin{equation} \label{tauperturb}
 \tau(z)=-i\frac{N_f}{2\pi}\log 
 \left(\frac z {\rho_L}\right) \,.
\end{equation}
The integration constant $\rho_L$ determines the string coupling.

The second equation of motion, Eq.~(\ref{eqnepsi}) is solved by
\begin{align}
e^{\psi(z, \bar z)} = \Omega \bar \Omega {\,\rm Im\,}(\tau) \,,
\end{align}
where $\Omega$ is an arbitrary holomorphic function. This function is
fixed by requiring modular invariance and regularity of the metric at
the location of the D7-branes. The function $e^{\Psi}$ is thus given
by
\begin{align}
e^{\Psi(z, \bar z)} = \tau_2(z) \vert 
\eta(\tau) \vert^4 \vert z \vert^{-{N_f}/{6}}
\end{align}
with $\eta$ the Dedekind eta function and $\tau_2={\rm Im}\, (\tau)$.
In the weak coupling region, $|z| \ll \rho_L$, $\tau_2$ is large and
the function $e^\Psi$ is well-approximated by
\begin{align}
e^{\Psi(z,\bar z)} \approx \tau_2 \,.
\end{align}
Asymptotically, for $|z| \gg \rho_L$, the complex structure $\tau$ as
given by Eq.~(\ref{jfu}) approaches the constant $\tau=j^{-1}(0)$ and
$e^{\Psi(z,\bar z)} \approx |z|^{-N_f/6}$. This leads to an
asymptotically flat metric with a deficit angle of $2\pi N_f/12$ as
can be seen from the transverse part of the D7 metric,
\begin{align}
ds^2_{\perp}= \rho^{-N_f/6} (d\rho^2 + \rho^2 d\varphi^2) = 
 \frac 1{(1-\frac{N_f}{12})^2} (d\rho'{}^2 + \rho'{}^2 (1-\frac{N_f}{12})^2 
d\varphi^2) \,,
\end{align}
where $\rho'=\rho^{1-N_f/12}$. This shows that the transverse space is
asymptotically conical for $N_f<12$ and asymptotically cylindrical for
$N_f=12$. For $N_f >12$ the transverse space is compact and in general
singular apart from the exceptional case $N_f=24$.

\section{Solving the equation (\ref{de3})} \label{appGP}
\setcounter{equation}{0}\setcounter{figure}{0}\setcounter{table}{0}

The second order differential equation (\ref{de3}) admits two independent 
solutions. One of them, which is well behaved near $x=0$, 
was found by Gesztesy and Pittner \cite{Gesztesy}:
\begin{align}
y(x) = \sum^\infty_{n=0} \lambda^n e^{2nx} p_{n} (x) \,, \label{GP}
\end{align}
where the polynomials $p_{n}(x)$ are defined by the recursion relation
\begin{align} \label{polynomials}
\left(4n^2+ 4n \frac{d}{dx} + \frac{d^2}{dx^2} \right) p_{n}(x)
&=x p_{n-1}(x) \,,\qquad p_{0}(x)=1\,.
\end{align} 
The series converges uniformly on the negative real line, $x < 0$.

Here we proceed to derive the other independent solution which is needed
in the construction of the warp factor of the backreacted D3/D7 geometry.
 
The recurrence relation 
\be
\bigg(4n^2+4n\frac{d}{d x} +\frac{d^2}{d x^2}\bigg)p_n(x)=x p_{n-1}(x)\,
\ee
can be decomposed into a set of algebraic equations for the expansion coefficients of the polynomials $p_n(x)$.
With the initial condition
\be
p_0(x)=a x+b\,,
\ee 
where $a,b$ are constants, we find that
\be
p_n(x)=\sum_{k=0}^{n+1}c^{(n)}_k x^k\,,
\ee
and where the coefficients $c^{(n)}_k$ satisfy the following system 
of equations
\bea
&&4n^2 c_{n+1}^{(n)}=c_{n}^{(n-1)} \,,\nn\\
&&4n^2 c_{n}^{(n)}+4n(n+1)c_{n+1}^{(n)}=c_{n-1}^{(n-1)}\,,\nn\\
&&4n^2 c_{n-k}^{(n)}+4n(n-k+1)c_{n-k+1}^{(n)}+(n-k+2)(n-k+1)c_{n-k+2}^{(n)}=
c_{n-k-1}^{(n-1)},\qquad k\geq 1\nn\\
&&4n^2c_0^{(n)}+4nc_{1}^{(n)}+2c_2^{(n)}=0\,.
\eea
According to our initial condition, we have $c_0^{(0)}=b, c_1^{(0)}=a$.
One can easily solve for the leading expansion coefficients 
(for $x\to-\infty$) 
\bea
c_{n+1}^{(n)}&=&\frac{a}{4^n (n!)^2},\qquad c_{n}^{(n)}=\frac{b}{4^n(n!)^2}-
\frac{a}{4^n(n!)^2}(n+\psi(n+1)+\gamma)
\nn\\
c_{n-1}^{(n)}&=&-\frac{1}{4n^2}\bigg(4n^2c_{n}^{(n)}+n(n+1)c_{n+1}^{(n)}+
\frac{1}{4(n-1)^2}\bigg(\nn\\&&4(n-1)^2c_{n-1}^{(n-1)}+n(n-1)c^{(n-1)}_n+
\frac{1}{4(n-2)^2}\bigg(\nn\\
&&4(n-2)^2c_{n-2}^{(n-2)}+(n-1)(n-2)c_{n-1}^{(n-2)}+
\dots\frac14\bigg(4c^{(1)}_1+1\cdot2 c_2^{(1)}\bigg)
\dots\bigg)\nn\\
&=&-\frac{1}{4^n(n!)^2}\bigg[bn-a\bigg(\frac{4n+3}4(\psi(n+1)+\gamma)+
\frac{n(2n-3)}4
\bigg)\bigg]\,,
\eea
where $\psi(n)=\frac{d\ln\Gamma(x)}{dx}$ is the polygamma function, and 
$\gamma=-\psi(1)$ is the Euler-Mascheroni constant.

Let us show that in the limit when the number of flavors is set to
zero (i.e.\ there are no D7-branes), we recover the modified Bessel
function $K_0(q\rho)$ which is needed to reconstruct the warp factor
of the geometry obtained by taking the near-horizon limit of a set of
overlapping D3-branes (i.e.\ the Green's function of the
six-dimensional flat space Laplace equation).

We begin by reintroducing the initial set of variables
\be
\lambda=-\frac{N_f}{2\pi}q^2 e^{4\pi/N_f},\qquad x=\ln\rho-\frac{2\pi}{N_f}
\ee
which yield
\bea
y(x)&=&\sum(\lambda e^{2x})^np_n(x)
\nn\\
&=&\sum \bigg(-\frac{N_f}{2\pi}\bigg)^n 
(q\rho)^{2n}\frac{1}{4^n(n!)^2}\bigg[a(\ln\rho-
\frac{2\pi}{N_f})^{n+1}\nn\\&&+\bigg(b-a\bigg(n+\gamma+\psi(n+1)\bigg)\bigg)
(\ln\rho-\frac{2\pi}{N_f})^n\nn\\&&+\bigg(bn-a(n+1)\bigg
(\frac{(n+1)^2-7}{6}+\gamma+
\psi(n+2)\bigg)\bigg)(\ln\rho-\frac{2\pi}{N_f})^{n-1}\nn\\&&+\dots\bigg]\,.
\eea

In order to obtain 
\be
K_0(q\rho)=\sum_{n=0}^\infty\frac{(q\rho)^{2n}(\psi(n+1)-\log(q\rho/2))}{
4^n(n!)^2}\, ,
\ee
we must choose
\be
b=\frac{2\pi a}{N_f}+a\gamma+a\ln(q/2)\,.
\label{barel}
\ee
Notice that with this choice we find that the only terms in the
$x(\rho)$ expansion of $y(x)$ that contribute in the limit $N_f\to 0$
are at most of order $x^{n-1}$, and at this order, the only
contribution comes from the $b$-dependent term.  The $an N_f^n x^n$
term is now canceled upon taking $N_f\to 0$ by the $bn N_f^n x^{n-1}$
term.  This completes the proof that $y(x(\rho))$ reduces indeed to
$K_0(q\rho)$ when setting the number of flavors to zero. Moreover, the
requirement that in this limit we obtain the modified Bessel function
$K_0(q\rho)$, allowed us to enforce the relation (\ref{barel}) between
the ab initio free parameters $a, b$. The remaining dependence on $a$
manifests as an overall proportionality coefficient which we fix by
the same requirement that in the limit $N_f\to 0$ the solution
$y(x(\rho))$ becomes $K_0(q\rho)$.  From now on, we set $a=-1$ and
(\ref{barel}).

The absolute value of the coefficients $c_k^{(n)}$ is bounded from above by   
\be
|c_{k}^{(n)}|\stackrel{<}{=}\frac{1}{4^n n!}\bigg(
|a|(\psi(n+1)+\gamma+1)C_{n+1}^{n+1-k} 2^{n+1-k}+
|b|C_{n}^{n-k}2^{n-k}\bigg)\,.
\ee
Consequently, the polynomials $p_{n}(x)$ obey the inequality
\be
|p_n(x)|\stackrel{<}{=}\frac{1}{4^n n!}\bigg(
|a|(\psi(n+1)+\gamma+1)(2+|x|)^{n+1}+|b|(2+|x|)^n\bigg)\,.
\ee
Therefore it can be shown that the series $y(x)=\sum \lambda^n
e^{2nx}p_n(x)$ is convergent given that each term in the series is
smaller than the corresponding term of a convergent series $\sum
|\lambda|^n e^{2nx}
\frac{1}{4^nn!}\bigg(|a|(\psi(n+1)+\gamma+1)(2+|x|)^{n+1}
+|b|(2+|x|)^n\bigg)$.  The convergence of the latter series is proven
by showing that the ratio of two consecutive terms
$a_{n}/a_{n-1}=\lambda e^x (2+x)
(|a|(2+|x|)(\psi(n)+\gamma+1)+|b|)/((a(2+|x|)(\psi(n-1)+\gamma+1)+|b|)4
n)$ approaches zero as $n\to\infty$.

\end{document}